\documentclass[a4page, aip,apl,amsmath,amssymb,reprint,superscriptaddress]{revtex4-1}
\usepackage{graphicx}
\usepackage{amsmath}
\usepackage{dcolumn}
\usepackage{bm}
\usepackage{bbold}
\usepackage{color}
\usepackage{amsfonts}
\usepackage{amssymb}
\usepackage{mathrsfs}
\usepackage{tabularx}
\usepackage{braket}
\usepackage{mathtools}
\usepackage{soul}			
\usepackage{booktabs}

\usepackage{caption}
\usepackage{subcaption}

\begin{document} 
\title{Electromagnetic properties of terbium gallium garnet at millikelvin temperatures and low photon energy}

\author{Nikita Kostylev}
\affiliation{ARC Centre of Excellence for Engineered Quantum Systems, School of Physics, University of Western Australia, 35 Stirling Highway, Crawley WA 6009, Australia}

\author{Maxim Goryachev}
\affiliation{ARC Centre of Excellence for Engineered Quantum Systems, School of Physics, University of Western Australia, 35 Stirling Highway, Crawley WA 6009, Australia}

\author{Pavel Bushev}
\affiliation{Experimentalphysik, Universit\"{a}t des Saarlandes, D-66123 Saarbr\"{u}cken, Germany}

\author{Michael E. Tobar}
\email{michael.tobar@uwa.edu.au}
\affiliation{ARC Centre of Excellence for Engineered Quantum Systems, School of Physics, University of Western Australia, 35 Stirling Highway, Crawley WA 6009, Australia}

\date{\today}


\begin{abstract}
Electromagnetic properties of single crystal terbium gallium garnet (TGG) are characterised from room down to millikelvin temperatures using the whispering gallery mode method. Microwave spectroscopy is performed at low powers equivalent to a few photons in energy and conducted as functions of magnetic field and temperature. A phase transition is detected close to the temperature of 3.5 Kelvin. This is observed for multiple whispering gallery modes causing an abrupt negative frequency shift and a change in transmission due to extra losses in the new phase caused by a change in complex magnetic susceptibility.
\end{abstract}

\maketitle

Quantum communication and information processing has experienced rapid growth in the recent years, instigated by significant advances in quantum technology. A lot of attention is currently drawn towards developing discrete elements of quantum circuits with a prospect of unifying them in a single Hybrid Quantum System (HQS)\cite{Stephens2013} in the future. One of the main complications, slowing down progress in this field of research, is the noticeable disconnect between the multitude of approaches in performing operations on quantum states and handling their storage and transfer. A typical issue is the frequency range each of these subsystems is limited to. For example, structures, such as superconducting qubits\cite{Xiang2013, Koch2007, Manucharyan2009}, or spins coupled to waveguides\cite{Burkard2006}, allow for processing and storage of information with particularly high fidelity, but operate primarily at microwave wavelengths. Long-range transfer of qubits, on the other hand, is realised best over optical links. In order to bridge the gap between these regimes, a device capable of performing high-efficiency, high-rate quantum frequency conversion is required. 

Translating quantum information between microwaves and optics is carried out using a system that has accessible energy states in both frequency domains. Suggestions have been made to use quantum mechanical resonators\cite{Stannigel2010, Andrews2014}, spin-doped dielectric crystals, such as iron ions in sapphire\cite{Farr2013}, Er:YSO\cite{Bushev2011,Probst2013, Williamson2014, Fernandez2015}, electro-optical effects in materials such as lithium niobate\cite{Rueda2016} and NV-centres in diamond\cite{Kubo2011}. The performance of these devices depends strongly on material properties, where of primary concern is the coherent interaction of system photons, and thus coupling strengths to both microwave and optical cavities play a major role. It has recently been shown that with certain ferrimagnetic crystals, such as yttrium iron garnet (YIG), it is possible to reach extreme coupling rates, including ultra-strong\cite{Goryachev2014} and superstrong coupling\cite{Kostylev2016}, where the system losses become comparable to mode frequency and free spectral range of the cavity respectively. Optical coupling to YIG, known for its good magneto-optical properties\cite{Cherepanov1993, Demokritov2008}, has also been realised, but was proven to be rather weak due to short absorption lengths\cite{Haigh2015}. This is quite a usual occurrence, as an operating approach in one frequency band typically introduces significant disadvantages in the other\cite{Goryachev2015}. Nevertheless, recent work has shown that bidirectional microwave-optical photon conversion is possible in YIG \cite{Hisatomi2016}, although the conversion efficiency in such a system was rather low, only on the order of $10^{-10}$. The search is therefore ongoing for magnetic materials that exhibit low losses at both sides of the spectrum, which may be mediated by spins and/or electronic properties and could result in high conversion rates. It is especially important to estimate these qualities at low temperatures and low excitation levels where quantum information may be preserved.

In this work, we characterize the electromagnetic properties of Terbium Gallium Garnet (TGG) single crystals at millikelvin temperatures and single photon levels by microwave spectroscopy. TGG, which has a chemical composition Tb$_3$Ga$_5$O$_{12}$ and a cubic garnet structure that forms a geometrically frustrated three dimensional Hyperkagone lattice\cite{Low2013}, is quite popular for its applications in optics\cite{Mahajan2013, Stevens2016, Yasuhara2014}. Similar to YIG, it has a high Verdet constant, resulting in a large magneto-optical effect, which increases substantially when the crystal is cooled towards cryogenic temperatures\cite{Majeed2013}. As a consequence, this crystal is considered particularly effective as a Faraday rotator and isolator in laser setups, especially due to its excellent thermal conductivity, low optical absorption coefficient and high laser-damage threshold\cite{Yin2015}. Recent work has shown interesting results with optical coupling to TGG using the inverse Faraday effect, exciting magnetic resonances at terrahertz frequencies\cite{Mikhaylovskiy2012, Mikhaylovskiy2013}. In these experiments, it has also been found that optical pumping with femtosecond laser pulses can be used to modify the magnetic state of Tb ions. Knowledge on accessing these states in the microwave regime could be potentially extremely beneficial in bridging the optical-microwave frequency gap for quantum conversion applications. Notwithstanding, except for a limited number of studies on magnetic properties, such as field-induced antiferromagnetism\cite{Kamazawa2008}, and discovery of anisotropic thermal conductivity, such as the phonon Hall effect\cite{Strohm2005}, dielectric properties of this material at low temperatures and microwave frequencies have not been thoroughly investigated.

\begin{figure}[ht!]
			\includegraphics[width=0.25\textwidth]{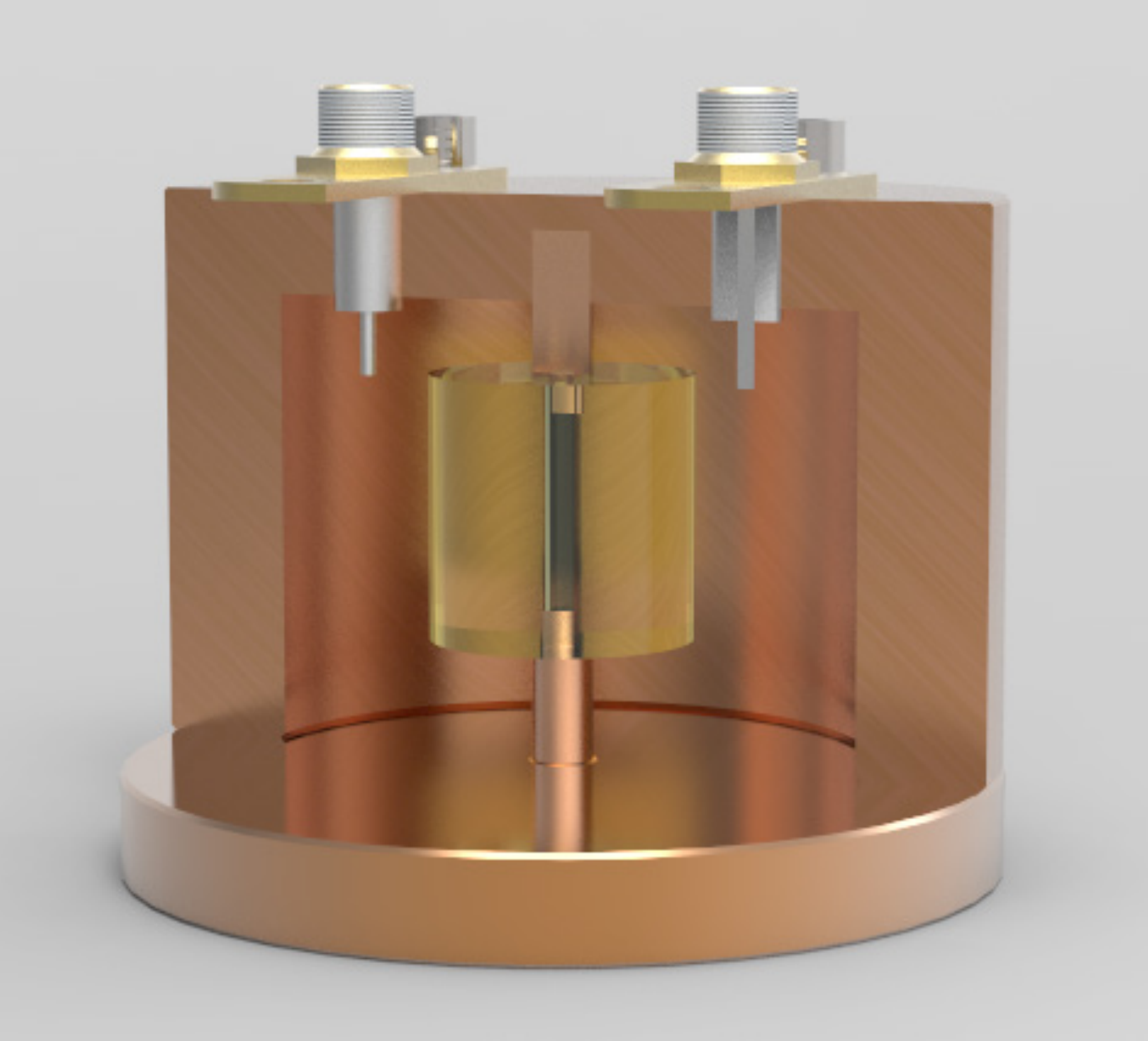}
	\caption{3D schematic showing the whispering gallery mode cavity formed by loading a cylindrical TGG crystal inside a cylindrical copper cavity.}
	\label{3dcad}
\end{figure}
\begin{figure}[ht!]
			\includegraphics[width=0.3\textwidth]{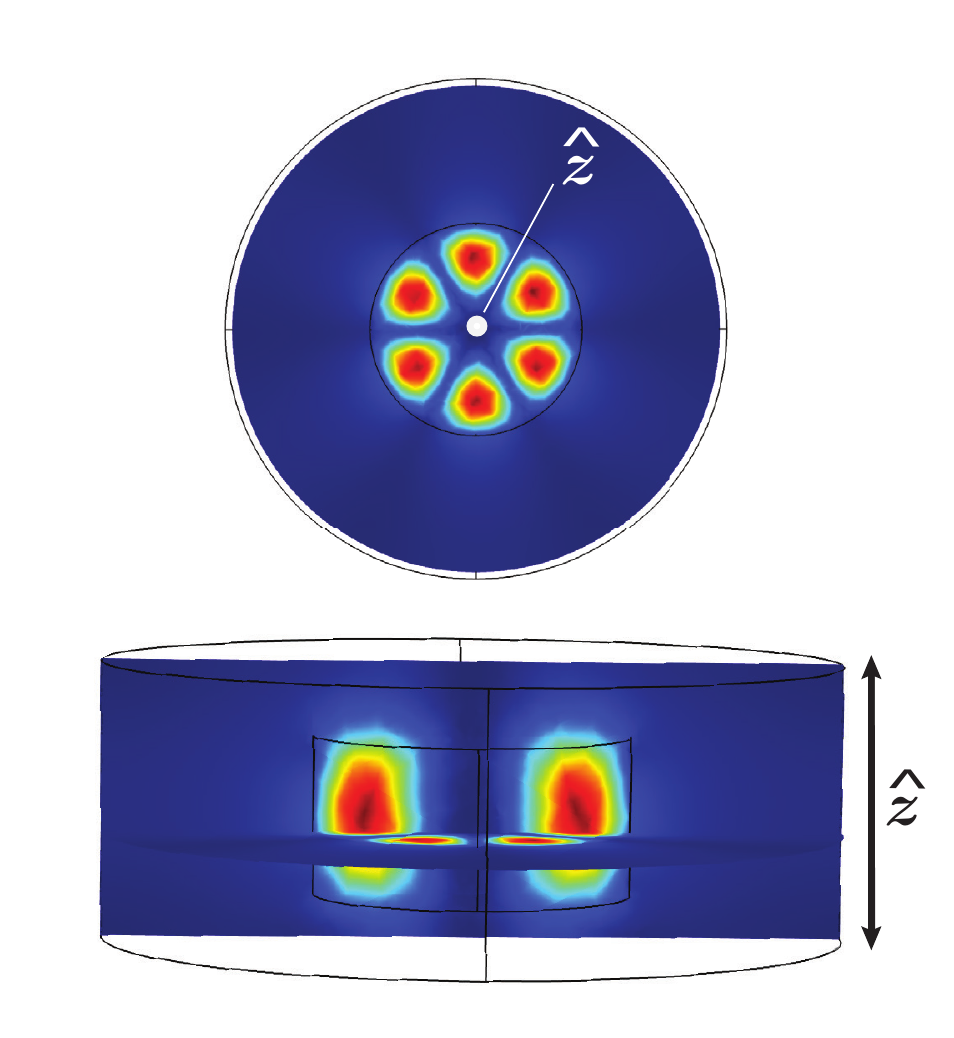}
	\caption{Simulated field density for a WGM with azimuthal mode number $m = 3$ and $\omega_0/2\pi$ = 8.84 GHz. }
	\label{wgm}
\end{figure}

As for the spectroscopy technique, the whispering gallery mode (WGM) approach was chosen. It is an extremely sensitive tool that has on many occasions been proven to generate accurate estimates of electromagnetic characteristics of materials\cite{Krupka1999} and identification of ion impurities\cite{Farr2015, Farr2013, Goryachev2014b}, including detection of paramagnetic spin flip transitions down to less than a few parts per trillion\cite{Kostylev2017a}. This method has also been used in discovering various nonlinear effects due to spin interactions, such as masing in a two-level system\cite{Bourhill2013} and four-wave mixing\cite{Creedon2012}.

To investigate this material, a cylindrical crystal of TGG was obtained with dimensions $15\times15$ mm (D$\times$H) and was mounted in a cylindrical cavity manufactured out of oxygen free (OFHC) copper to realise a dielectric loaded WGM resonator. The sample was undoped, with only traces of natural impurities expected to be present in the structure. The crystal was fixed in place by thermally conductive copper mounts, inserted into a 1mm diameter bore in the middle of the crystal (Fig.~\ref{3dcad}). The cavity resonator assembly was thermally connected to a 20mK stage of a dilution cryostat and subjected to a magnetic field of up to 7T in strength, generated by a superconducting magnet. The sample was tested using a vector network analyzer in transmission, with the room temperature signal level reduced by a series of cryogenic attenuators (-40dB in total) and the output amplified with two 20dB low noise amplifiers with one thermalised to 4K and the other to room temperature. The field was excited using straight microwave antennae, coupling primarily to the WGH modes (classified as having the dominant \textbf{B}-field aligned along the $z$-axis of the cylinder and \textbf{E}-field in the radial direction) in the cavity. An example of the WGM is given in Fig.~\ref{wgm}.

\begin{figure}[ht!]
			\includegraphics[width=0.45\textwidth]{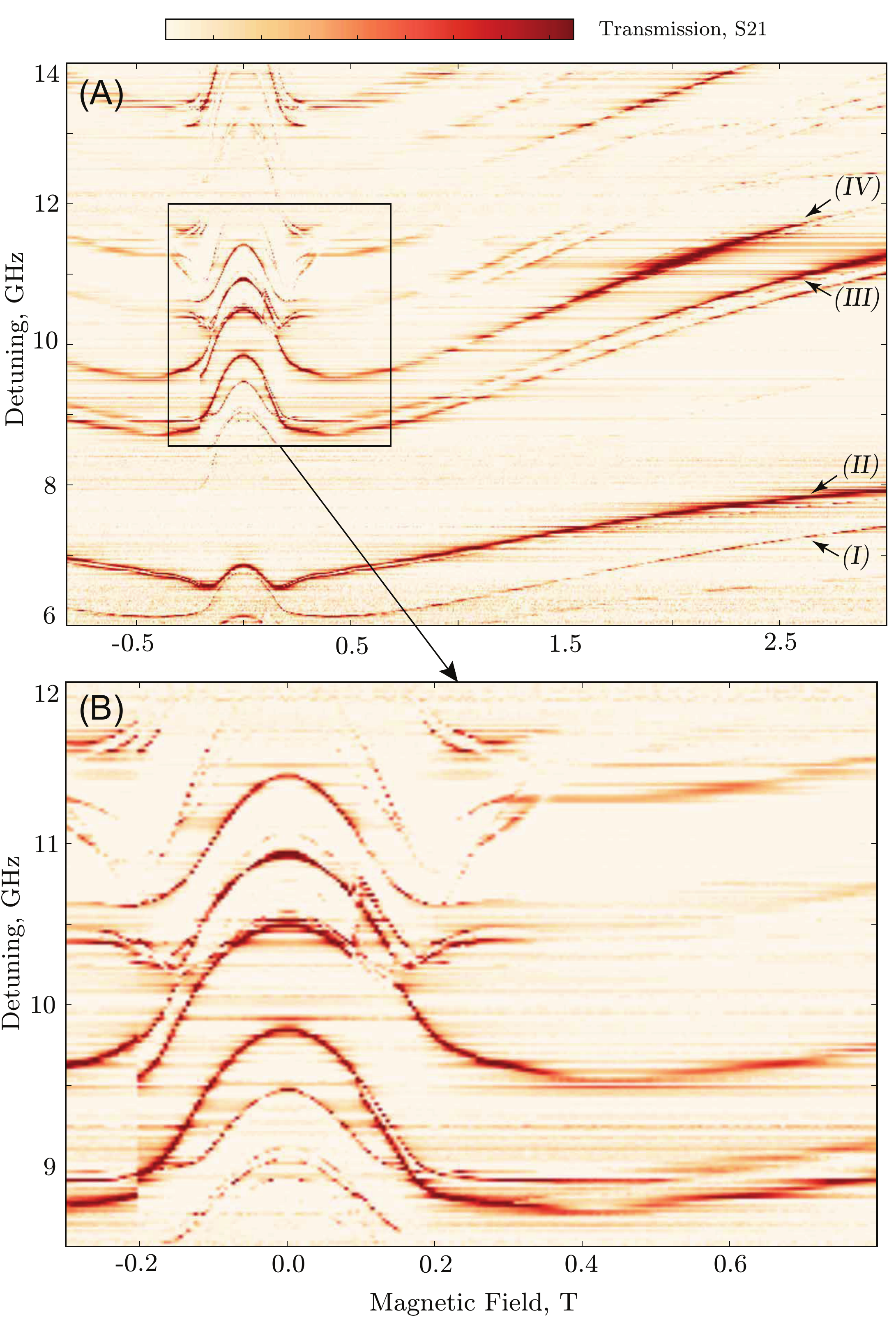}
	\caption{Whispering gallery mode frequency response of the TGG crystal as a function of the external magnetic field. A section of the plot (A), showing the low applied magnetic field behaviour, is magnified in (B) to reveal the complex level crossing structure. The most prominent modes are labeled with indexes (I-IV) and will be referred to in latter figures.}
	\label{bfield}
\end{figure}

\begin{figure}[ht!]
			\includegraphics[width=0.45\textwidth]{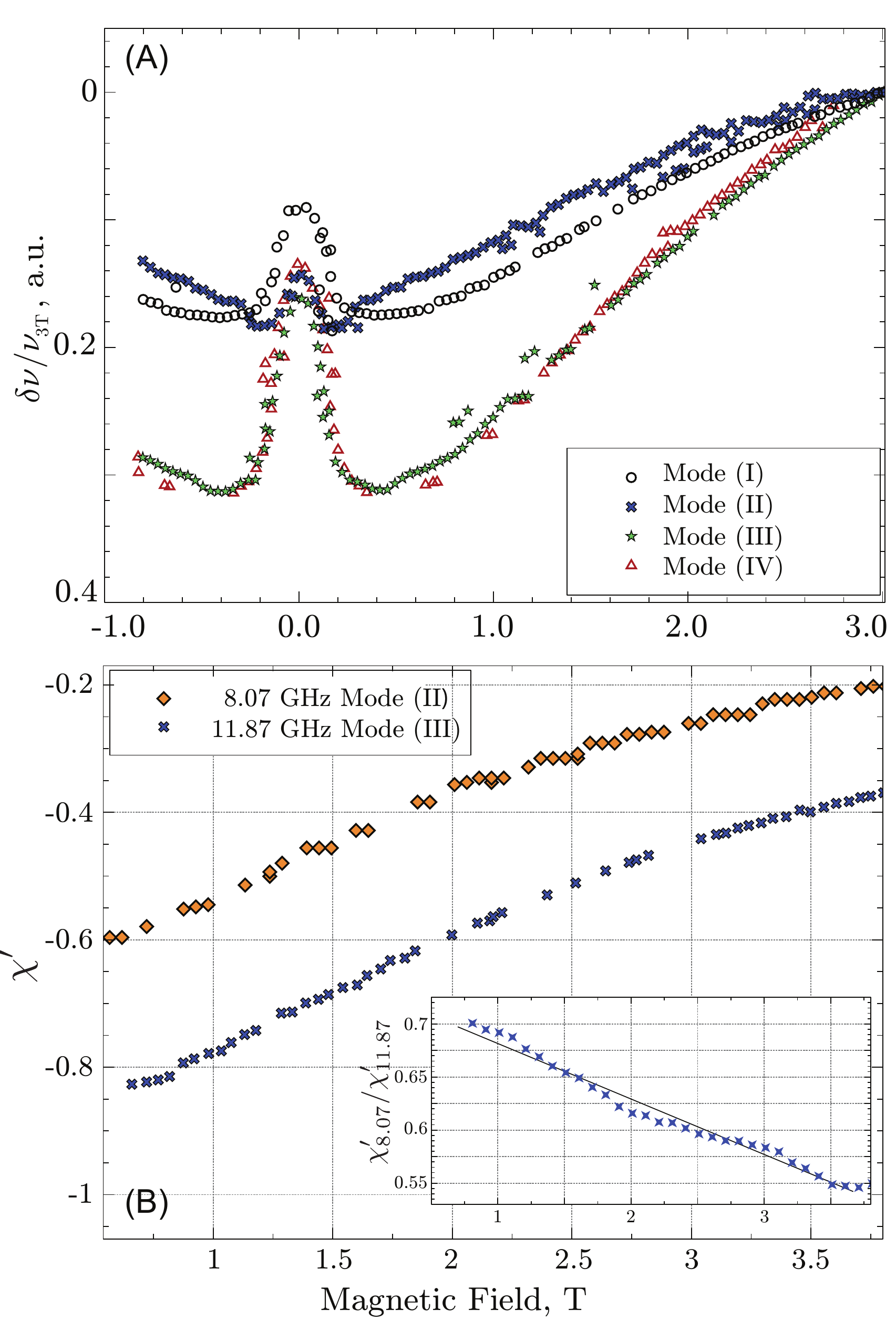}
	\caption{(A) Normalized frequency as a function of the external magnetic field for selected WGM modes. (B) Real part of complex magnetic susceptibility as a function of external magnetic field for selected WGM modes. The inset shows the ratio between the two susceptibility curves.}
	\label{chi}
\end{figure}

The experimental data for the magnetic field sweep as a function of frequency detuning at the temperature of 20mK are shown in Fig.~\ref{bfield} and Fig.~\ref{chi}. A complex interaction structure is present in the vicinity of zero applied field, with the mode frequencies experiencing a rapid repulsion as the field strength increases. The near zero region at these ultra-low temperatures is characterised by the Tb$^{3+}$ ion ensemble being in the antiferromagnetic phase due to the significance of the spin-spin interaction between Tb$^{3+}$ ions\cite{Low2013}. Very high magnetic field tunability reaching $100$GHz/T can be seen in the small-scale interaction regions around $B = \pm 0.15$T. At higher fields, the Zeeman term prevails and the ensemble is effectively characterised by the paramagnetic phase. Also, the response is mirror symmetric around the zero field with no significant dependence on the measurement history, i.e. implying no detectable hysteresis. 

Figure~\ref{chi}(A) compares frequency deviations of different modes (given by $\delta \nu = \nu - \nu_{3T}$) tuned by the magnetic field scaled by their frequencies at 3T. Although all curves have the same structure with three turning points, the result demonstrates that the magnitude of the frequency deviation grows with the resonance frequency. The effect is similar to the one recently observed in a ferromagnetic Erbium-doped YAG crystal\cite{Farr2015}. 

The shifts in the WGM frequencies can be related to the complex magnetic susceptibility of the material \cite{Creedon2010}:

\begin{equation}
\label{eqn:Chi}
\chi' = 2 \frac{\delta\nu}{p_{m \perp} \nu_0},
\end{equation}

where $p_{m \perp}$ is the perpendicular magnetic filling factor and $\delta \nu$ is the frequency deviation of the mode in relation to  the resonance frequency $\nu_0$, as determined for high $B$ fields. Fig.~\ref{chi}(B) shows the real part of the magnetic susceptibility $\chi'$ for a wide range of magnetic fields. A frequency-dependent, dispersive response can be observed when the material is in the effectively paramagnetic phase, with an approximately linear relationship between $\chi'$ values of the most prominent modes, as highlighted in the figure's inset.

\begin{figure}[ht!]
			\includegraphics[width=0.5\textwidth]{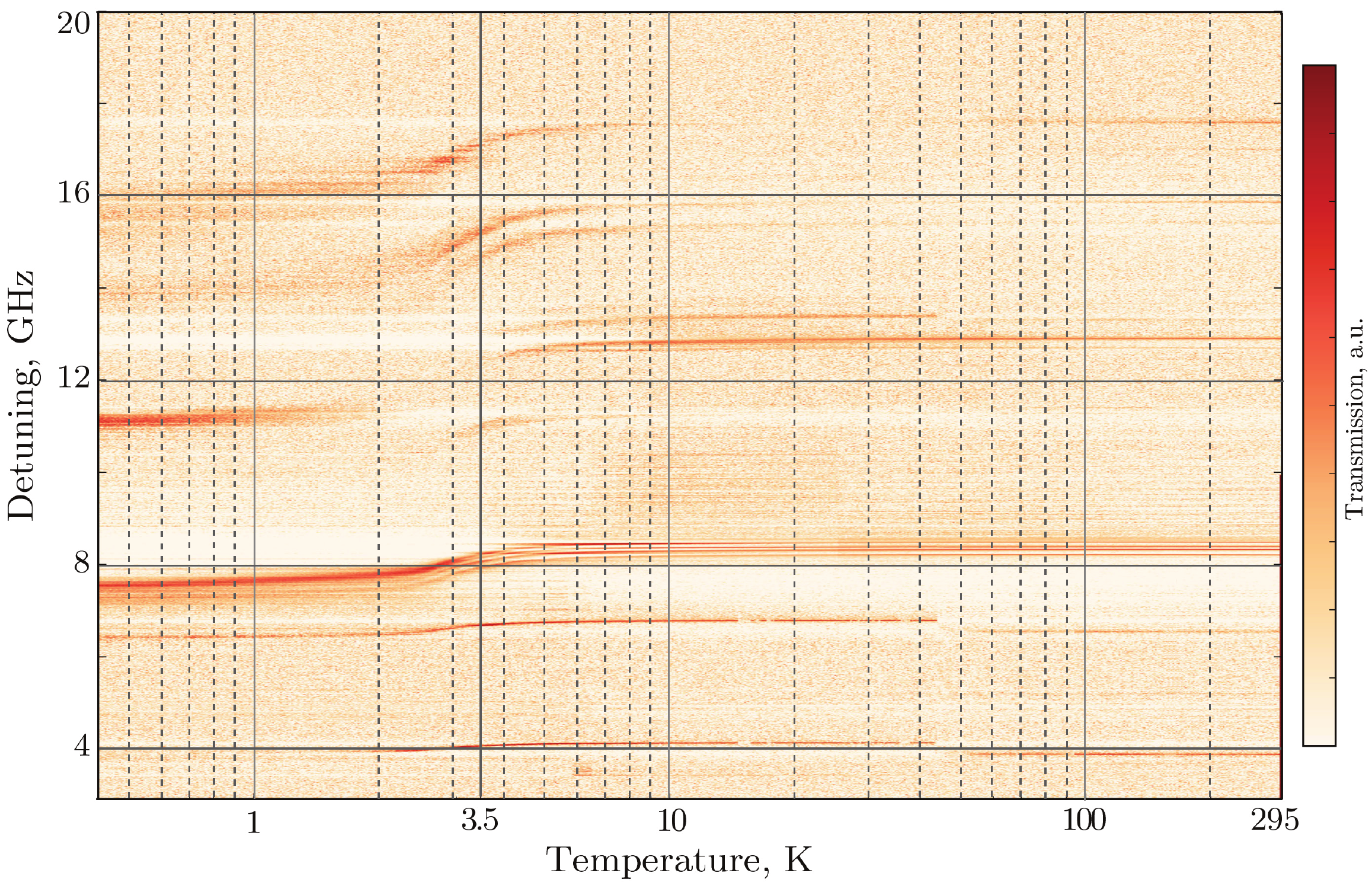}
	\caption{Density plot of cavity transmission for the temperature range of 0.02-295K. A steep resonance frequency shift can be observed for the WGM modes over a wide range of frequencies, occurring in the vicinity of the phase transition around 3.5K. }
	\label{temp_density}
\end{figure}

\begin{figure}[ht!]
			\includegraphics[width=0.45\textwidth]{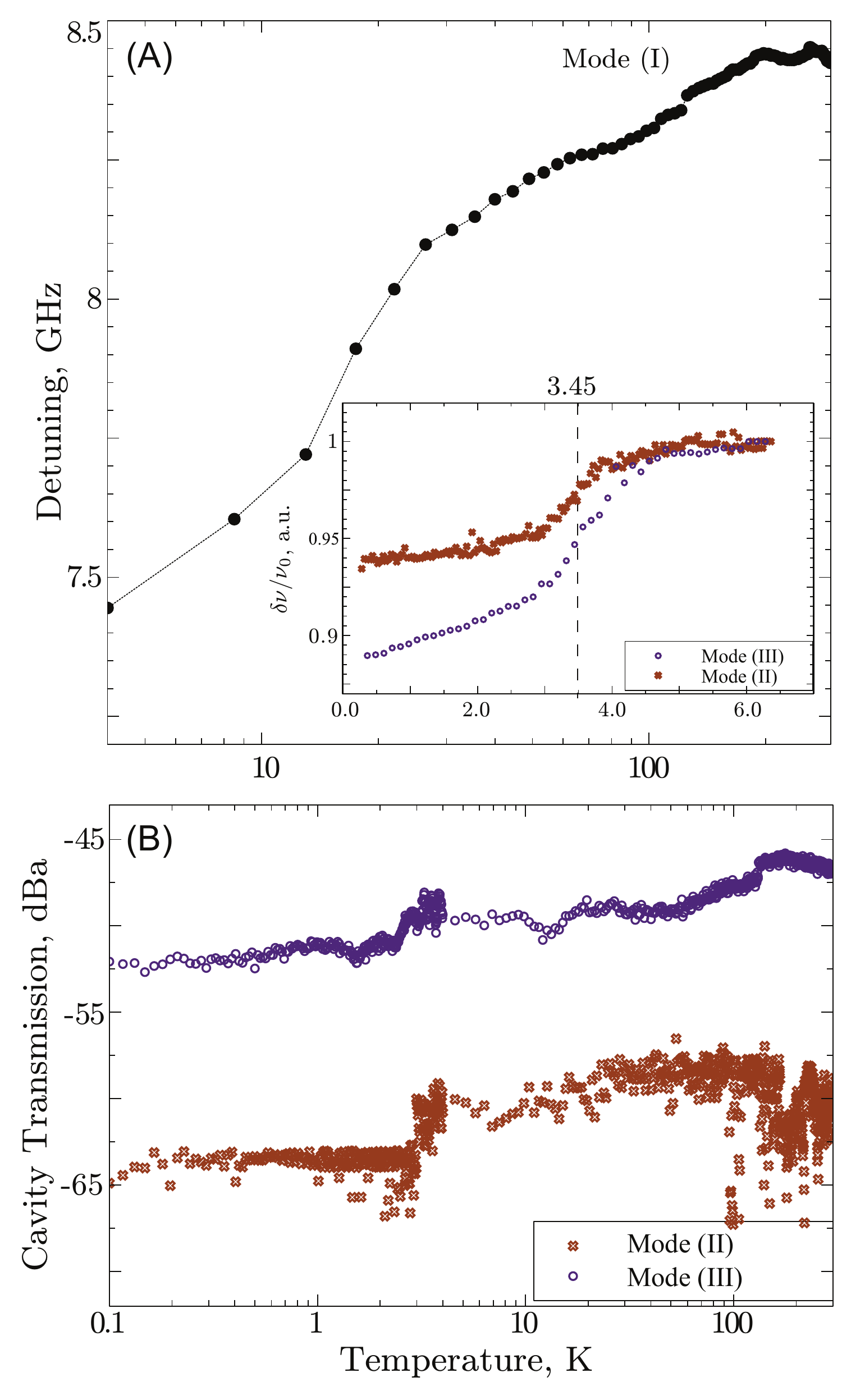}
	\caption{(A) Frequency response of the 8.42GHz mode for the temperature range of 0.02-295K. The inset shows the normalized resonant frequency shift as a function of temperature for selected WGM modes in the vicinity of the phase transition.  (B) Cavity transmission for selected WGM modes for a wide temperature range, showing increased loss through the phase transition. The plots indicate a significant increase in transmission and frequency detuning as the material undergoes a phase change at 3.5K.}
	\label{temp}
\end{figure}

Figures~\ref{temp_density} and \ref{temp} show the resonant frequency behaviour of the modes as a function of temperature, measured as the sample was being cooled down. At around 3.5K, a clear shift in all mode frequencies can be observed. This appears to be the result of a phase transition, from the paramagnetic to the antiferromagnetic phase. This temperature is higher than that for the reports of a N\'{e}el state, onset of which has been observed at 0.35K previously\cite{Kamazawa2008}. The modes presented in Figure~\ref{temp} have been identified using finite element analysis as fundamental WGH modes of a low azimuthal number and are given in Table~\ref{table1}. The values for the frequency shift $\delta \nu$ due to the phase transition are also presented in the table along with the estimates of the Quality factors $Q$ for both low and high temperature regimes. The measurements demonstrate an increase in losses in the system, which can also be observed in the transmission profiles of the modes, shown in Figure~\ref{temp}(B).

\begin{table}[]
\centering
\caption{Change in WGM resonance,  including frequency shift $\Delta \nu$, quality factors $Q_\text{room}$ and $Q_\text{mK}$ and average photon number $N_\text{Ph}$ for selected modes as the TGG crystal is cooled down to cryogenic temperatures, experiencing a phase transition.}
\label{table1}
\begin{tabular}{cccccc}
\hline
$\nu$ (GHz) & Mode    & $\delta \nu$ (MHz) & $Q_\text{room}$ & $Q_\text{mK}$ & $N_\text{Ph}$ \\ \hline
\multicolumn{1}{|c|}{6.67}      & \multicolumn{1}{c|}{WGH$_\text{2,0,0}$} & \multicolumn{1}{c|}{354}         & \multicolumn{1}{c|}{238}        & \multicolumn{1}{c|}{138}      & \multicolumn{1}{c|}{29.8} \\ \hline
\multicolumn{1}{|c|}{8.84}      & \multicolumn{1}{c|}{WGH$_\text{3,0,0}$} & \multicolumn{1}{c|}{912}         & \multicolumn{1}{c|}{533}        & \multicolumn{1}{c|}{210}      & \multicolumn{1}{c|}{25.8} \\ \hline
\multicolumn{1}{|c|}{13.87}     & \multicolumn{1}{c|}{WGH$_\text{5,0,0}$} & \multicolumn{1}{c|}{885}         & \multicolumn{1}{c|}{348}        & \multicolumn{1}{c|}{118}      & \multicolumn{1}{c|}{5.9}  \\ \hline
\end{tabular}
\end{table}

In conclusion, we utilised the multi-mode whispering gallery mode technique to analyse the electromagnetic properties of single crystal terbium gallium garnet from room temperature to millikelvin temperatures.  We observed a phase transition at around 3.5K, which caused an abrupt negative frequency shift and a change in transmission due to extra losses in the new phase due to the change in complex magnetic susceptibility.\\

This work was supported by Australian Research Council grant CE110001013. We thank Warrick Farr for assistance with data acquisition.

\bibliography{biblio}

\end{document}